\documentclass[12pt,preprint2]{aastex}
\newcommand{\Fermi}{\emph{Fermi}}
\newcommand{\g}{$\gamma$}
\newcommand{\lambar}{\lambda\llap {--}}

\usepackage{lineno}
\usepackage{fancyhdr}
\begin{document}
\title{Very Fine Time-Resolved Spectral Studies of the Vela Pulsar with the \Fermi{} Large Area Telescope}
\author{T.~J.~Johnson\altaffilmark{1,2}, \"O.~\c{C}elik\altaffilmark{3,4}, M. Kerr\altaffilmark{5} A.~K.~Harding \altaffilmark{2}, G.~A.~Caliandro \altaffilmark{6}, on behalf of the \Fermi{} LAT Collaboration and the Pulsar Timing Consortium}
\altaffiltext{1}{Department of Physics, University of Maryland, College Park, MD 20742, USA; Tyrel.J.Johnson@nasa.gov}
\altaffiltext{2}{NASA Goddard Space Flight Center, Greenbelt, MD 20771, USA}
\altaffiltext{3}{Center for Research and Exploration in Space Science and Technology (CRESST) and NASA Goddard Space Flight Center, Greenbelt, MD 20771, USA}
\altaffiltext{4}{Department of Physics and Center for Space Sciences and Technology, University of Maryland Baltimore County, Baltimore, MD 21250, USA}
\altaffiltext{5}{Department of Physics, University of Washinton, Seattle, WA 98195-1560, USA}
\altaffiltext{6}{Institut de Ciencies de l'Espai (IEED-CSIC), Campus UAB, 08193 Barcelona, Spain}

\begin{abstract}
The Vela pulsar is one of the most exciting \g-ray sources and has been at the forefront of high-energy pulsar science since the detection of \g-ray pulsations at the radio period by SAS-2 in 1975. With the unprecedented angular resolution, effective area, field of view, and timing resolution, in the GeV band, of the Large Area Telescope (LAT) on the \emph{Fermi Gamma-ray Space Telescope}, the light curve of the Vela pulsar can be studied in greater detail than ever before. Using a timing solution derived solely from the LAT data, phase aligned with the radio emission, the spectrum of the Vela pulsar has been fit in intervals as small as 0.0016 in phase. Significant variation is seen in the cutoff energy and photon index across the light curve, strongly supporting curvature radiation as the source of the high-energy \g-rays from the Vela pulsar.
\end{abstract}

\section{INTRODUCTION}
The Vela pulsar has a long history of \g-ray observations dating back to the point source discovery and pulsed detection with SAS-2 (Thompson et al.1974 and Thompson et al.1975, respectively).  Phase-resolved spectral results followed using data from COS-B \citep{gren} and EGRET (Kanbach et al.1994 and Fierro et al.1998) suggesting significant variation of the spectrum across the pulse, with the hardest emission observed between the two main \g-ray peaks.

With the advent of the \emph{Fermi Gamma-ray Space Telescope} (\Fermi{}), it is now possible to study the \g-ray emission from the Vela pulsar in unprecedented detail.  The main instrument aboard \Fermi{} is the Large Area Telescope (LAT) \citep{LAT}.  Using early calibration data \citep{abdoa} and $\sim$2 months of survey data the LAT produced the highest resolution high-energy (HE, $\geq0.1$ GeV) light curve of the Vela pulsar to date, identified a third peak which moves with increasing energy, and ruled out low-altitude magnetospheric emission models \citep{abdob}.  With 11 months of survey data it is now possible to perform the finest phase-resolved study of the Vela pulsar to date by fitting the spectrum in 101 variable-width bins across the pulse. 

\section{OBSERVATIONS}

\subsection{Timing}
LAT data was used to construct a timing solution for the Vela pulsar with 63 $\mu$s residuals using techniques described in \citet{ray}.  The \g-ray timing solution\footnote{The timing solution will be made available through the Fermi Science Support Center http://fermi.gsfc.nasa.gov/ssc/data/access/lat/\\ephems/} was phase aligned with data from the Parkes Radio Telescope \citep{manch}.

\begin{figure*}[Ht]
\epsscale{1.5}
\plotone{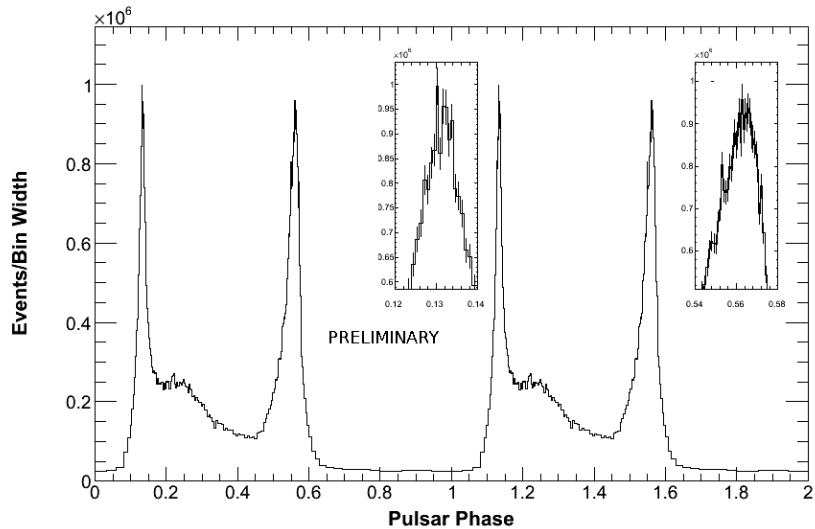}
\caption{Folded light curve of the Vela pulsar for events with reconstructed energies $\geq0.02$ GeV and within max[$1.3^{\circ}$,$1.6^{\circ}-3*\rm{log}_{10}(\rm{E}/1\ \rm{GeV})$] of the radio position, each bin has 750 counts.}
\end{figure*}

\subsection{Data Selection}
LAT event data from 4 August 2008 to 4 July 2009 with reconstructed energies from 0.02 to 100 GeV and belonging to the ``Diffuse" class \citep{LAT}, as defined under the P6\_V3 instrument response functions, were selected from within $15^{\circ}$ of the radio position.  Time periods when the limb of the Earth infringed upon the region of interest and when the rocking angle of the LAT exceeded $43^{\circ}$ were excluded to reduce background.  Events were phase-folded using the Fermi plug-in now provided with the TEMPO2\footnote{http://tempo2.sourceforge.net/} software \citep{hobbs}.  The resulting \g-ray light curve is shown in Figure 1, a detailed analysis of the light curve can be found in \citet{abdoc}.

\section{SPECTRAL ANALYSIS}

\subsection{Phase-Averaged}
Events with energies $\geq0.1$ GeV were used for spectral analysis.  The Fermi Science Tools\footnote{http://fermi.gsfc.nasa.gov/ssc/data/analysis/scitools/\\overview.html} (STs) v9r15p2 were used to perform a binned maximum likelihood analysis (Cash 1979 and Mattox et al. 1996) modeling all point sources found above the background with a test statistic $\geq25$ in a preliminary version of the 1FGL catalog \citep{abdod}; an extended source at the position of the Vela X pulsar wind nebula (PWN) \citep{abdoe}; and the v02 diffuse backgrounds, included with the Fermi STs.
The phase-averaged spectrum, Figure 2, is best fit as an exponentially cutoff power law, Equation 1, with the $b$ parameter $<$ 1:
\begin{equation}
\frac{dN}{dE} = \rm{N}_{0} \Big(\frac{\rm{E}}{1\ \rm{GeV}}\Big)^{-\Gamma} \exp\Big[ -\Big(\frac{\rm{E}}{\rm{E}_{C}}\Big)^{b} \Big]\ ,
\end{equation}
likely due to the superposition of many spectral components with $b\ \equiv\ 1$ and varying values of E$_{C}$ and $\Gamma$ through the pulse.  The best-fit spectral parameters are given in Table 1.

\begin{figure}[h]
\plotfiddle{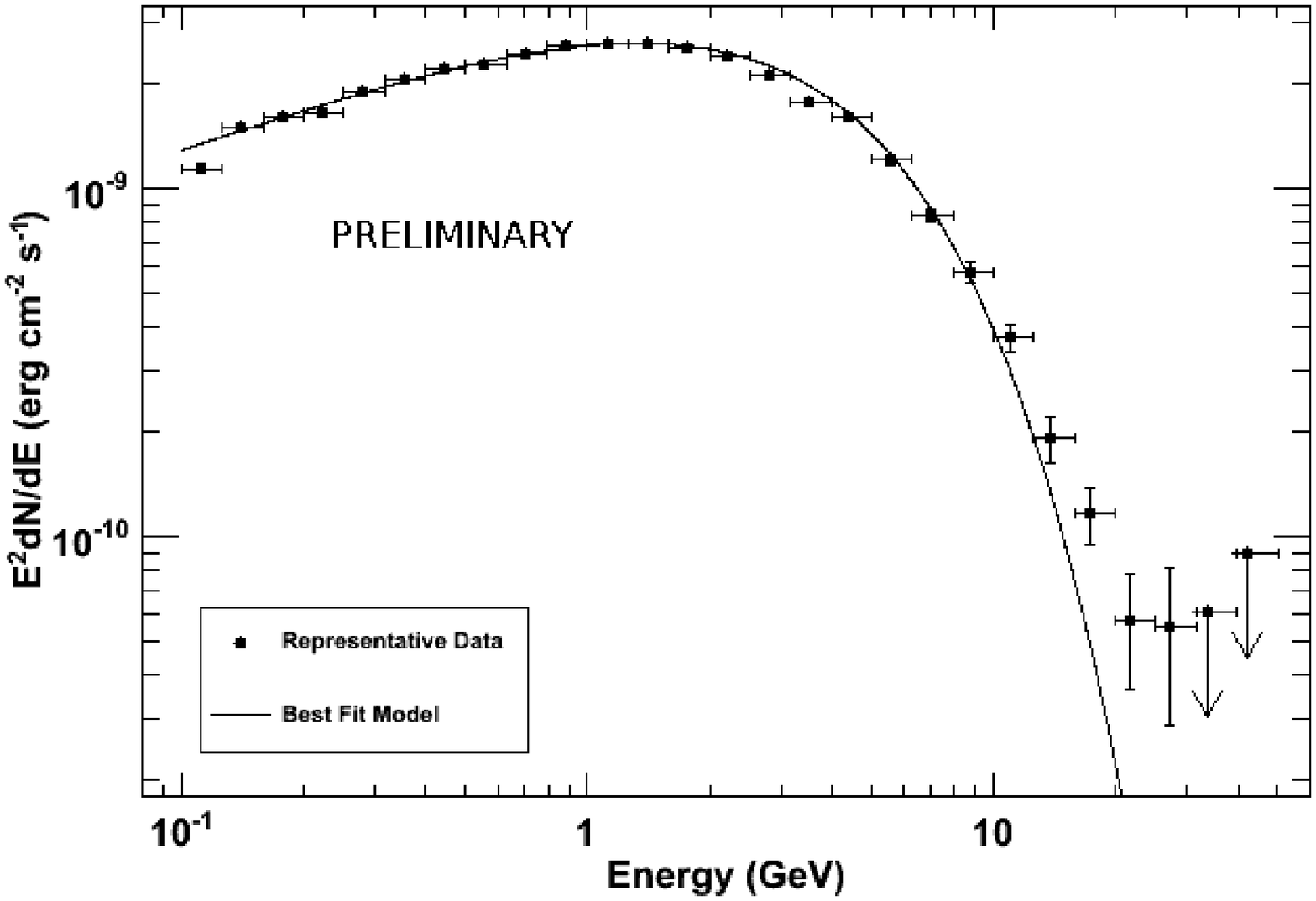}{0pt}{0}{270}{192.9}{-18}{0}
\caption{Phase-averaged \g-ray spectrum of the Vela pulsar.  Best-fit model with $b\ <\ 1$.  The representative data points are from likelihood fits to each energy range with the pulsar spectrum modeled as a power law.}
\end{figure}

\begin{deluxetable}{|l|r|}
\tablewidth{0pt}
\tablecaption{Vela Phase-Averaged Spectral Parameters\tablenotemark{a}}
\startdata
\hline \rm{N}$_{0}$ ($10^{-6}\ \rm{cm}^{-2}\ \rm{s}^{-1}\ \rm{GeV}^{-1}$) & $3.73\pm0.31\pm1.04$\\
\hline $\Gamma$ & $1.37\pm0.03^{+0.07}_{-0.03}$\\
\hline E$_{C}$ (GeV) & $1.31\pm0.18^{+1.0}_{-0.5}$\\
\hline $b$ & $0.68\pm0.03^{+0.18}_{-0.10}$\\
\hline Flux (0.1-100 GeV) ($10^{-5}\ \rm{cm}^{-2}\ \rm{s}^{-1}$) & $1.07\pm0.01\pm0.03$\\
\enddata
\tablenotetext{a}{First errors are statistical, second are systematic.}
\end{deluxetable}

\subsection{Phase-Resolved}
The light curve was divided into 101 variable-width bins with 1500 events each ($\geq0.1$ GeV) using the energy-dependent angular selection described in the caption of Figure 1.  The smallest bin has a width of 0.0016 in phase ($\sim142\ \mu$s).  The spectrum of the Vela pulsar was fit assuming a simple exponentially cutoff power law, $b\ \equiv\ 1$, in each phase bin.  Normalizations of the diffuse backgrounds and point sources within $5^{\circ}$ of the pulsar were also kept free.  The observed phase trends in $\Gamma$ and E$_{C}$ are shown in Figures 3 and 4, respectively.

\begin{figure}[h]
\plotfiddle{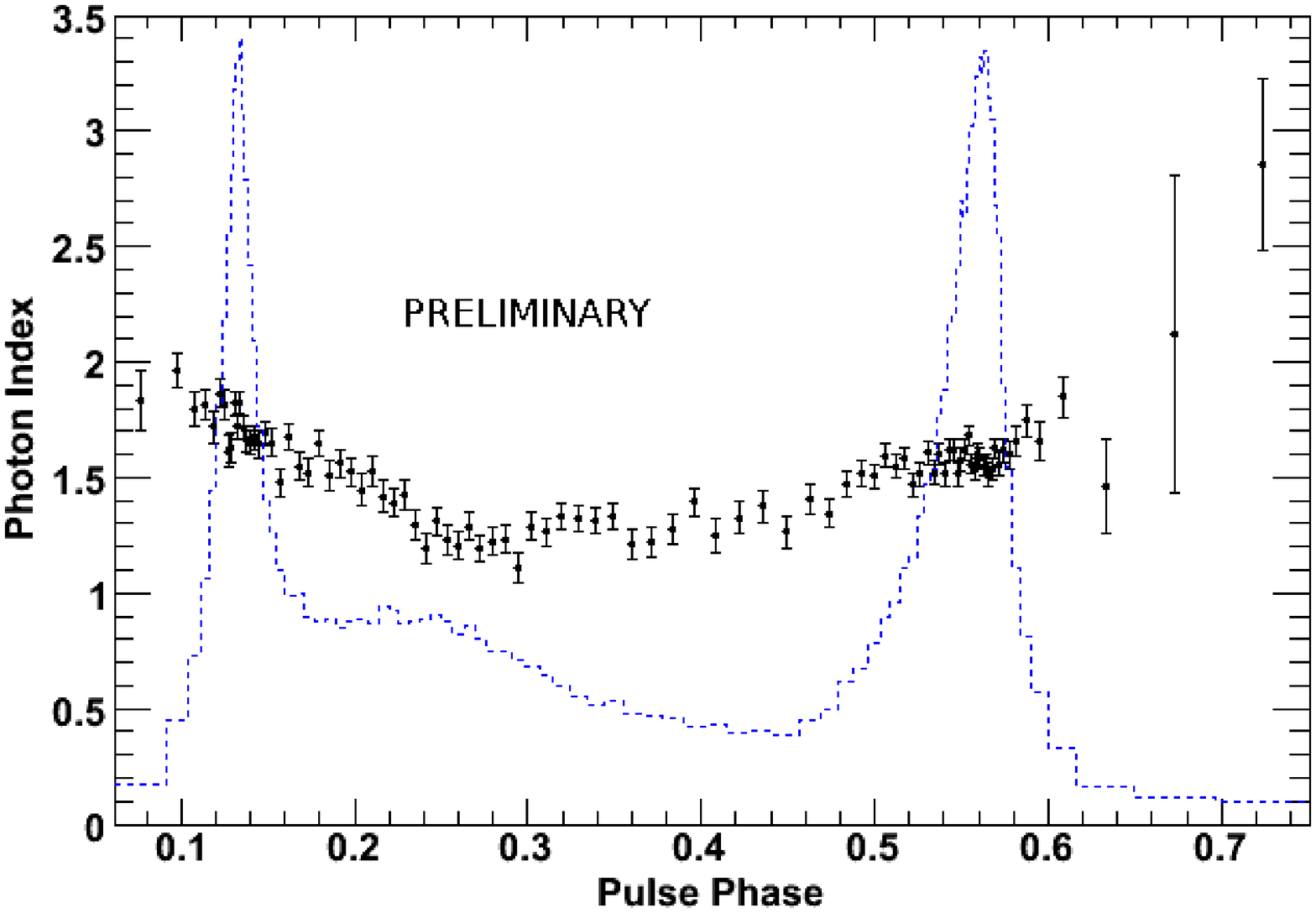}{0pt}{0}{270}{192.9}{-18}{0}
\caption{Photon index vs. phase. Errors are statistical only.  Results are only shown for bins in which the pulsar was found abve the background with a test statistic $\geq25$.}
\end{figure}

Significant variation is seen in both parameters, confirming that the hardest emission is between the two peaks.  E$_{C}$ rises sharply through the main peaks and, suprisingly, between the peaks as well, following the position of the third peak observed at higher energies.  Phase-resolved analyses of EGRET data suggested a drastic change in $\Gamma$ through the two main peaks; however, the LAT data does not confirm this, instead finding $\Gamma$ to be very consistent with a constant value through both peaks.

\begin{figure}[h]
\plotfiddle{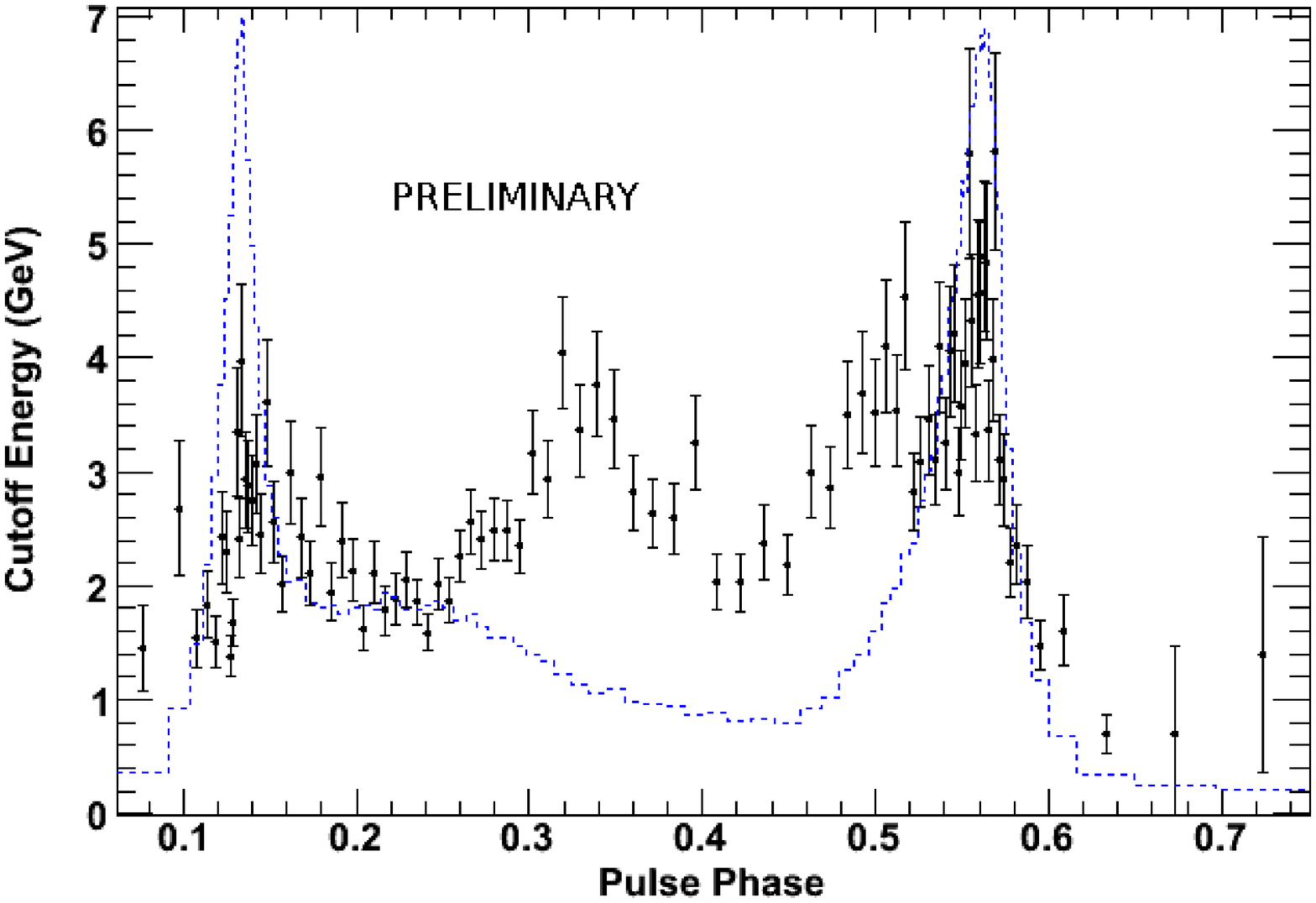}{0pt}{0}{270}{192.9}{-18}{0}
\caption{Cutoff energy vs. phase. Errors are statistical only.  Results are only shown for bins in which the pulsar was found above the background with a test statistic $\geq25$.}
\end{figure}

To evaluate the significance of the features in Figures 3 and 4, the pulsar and surrounding region were simulated using the Fermi ST \emph{gtobssim} and the built-in \emph{PulsarSpectrum} \citep{raz}.  The LAT-only timing parameters and spectral parameters from a phase-averaged fit with $b\ \equiv\ 1$ were used as input to the simulation.  The Vela X PWN was included as a point source.  The simulation suggests that point-to-point variations of 0.6 GeV in E$_{C}$ and 0.05 in $\Gamma$ should be expected from the fitting technique.  As such, point-to-point variations less than these values can not be considered significant.

\begin{figure}[h]
\plotfiddle{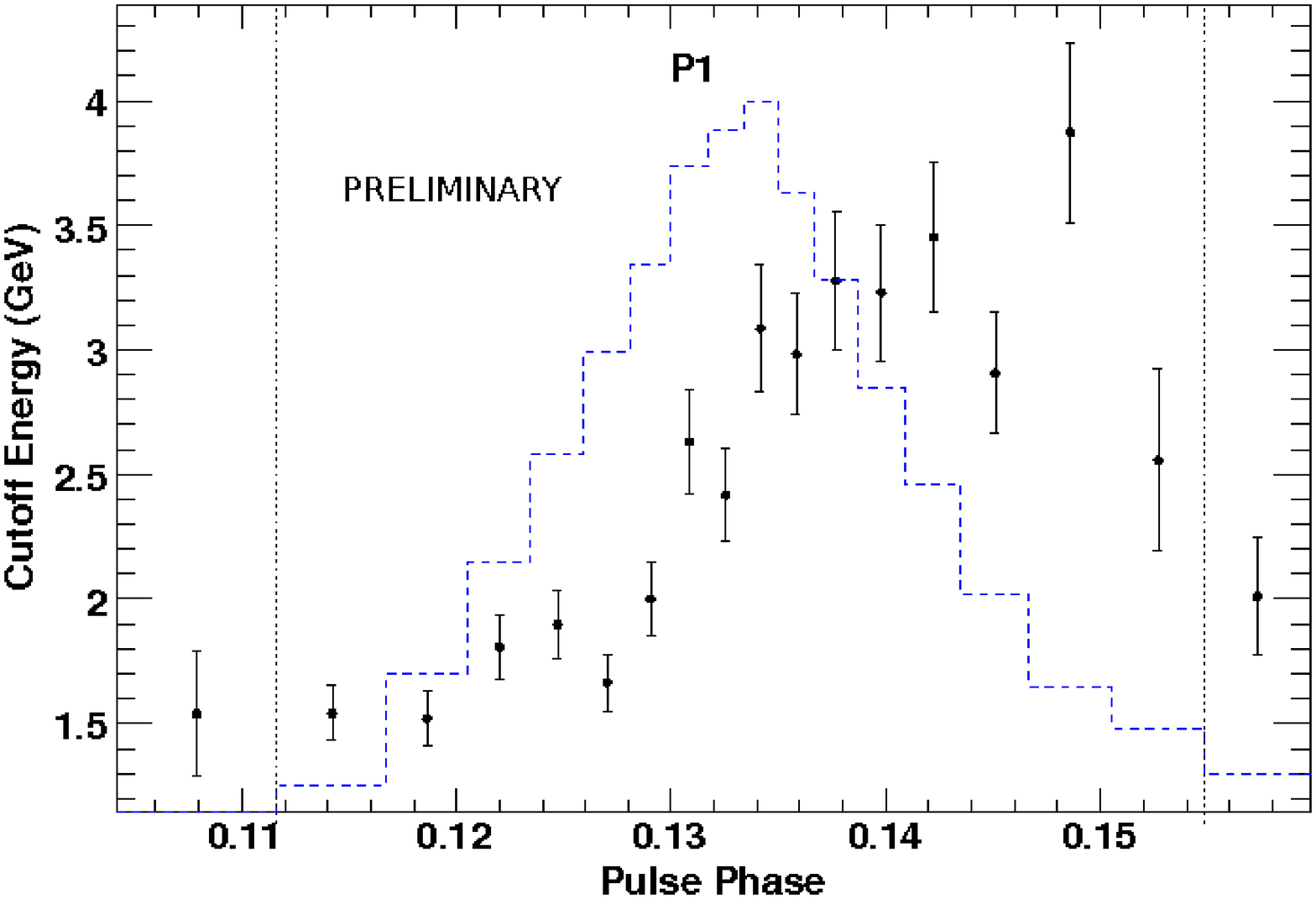}{0pt}{0}{270}{192.9}{-18}{0}
\caption{Cutoff energy vs phase for the first peak.  The photon index was held fixed to the best fit value of $1.72\pm0.01$ between the dashed vertical lines which correspond to the phase range $0.112\leq\phi\leq0.155$.}
\end{figure}

To better evaluate the behavior of E$_{C}$ in the two main peaks the fit was repeated but with $\Gamma$ fixed to the best-fit values of $1.72\pm0.01$ and $1.58\pm0.01$ for the first and second peaks, respectively.  The resulting trends in E$_{C}$ are shown in Figures 5 and 6 for the first and second peaks, respectively.  E$_{C}$ rises fairly smoothly in both peaks with maxima near the start of the trailing edges.

\begin{figure}[h]
\plotfiddle{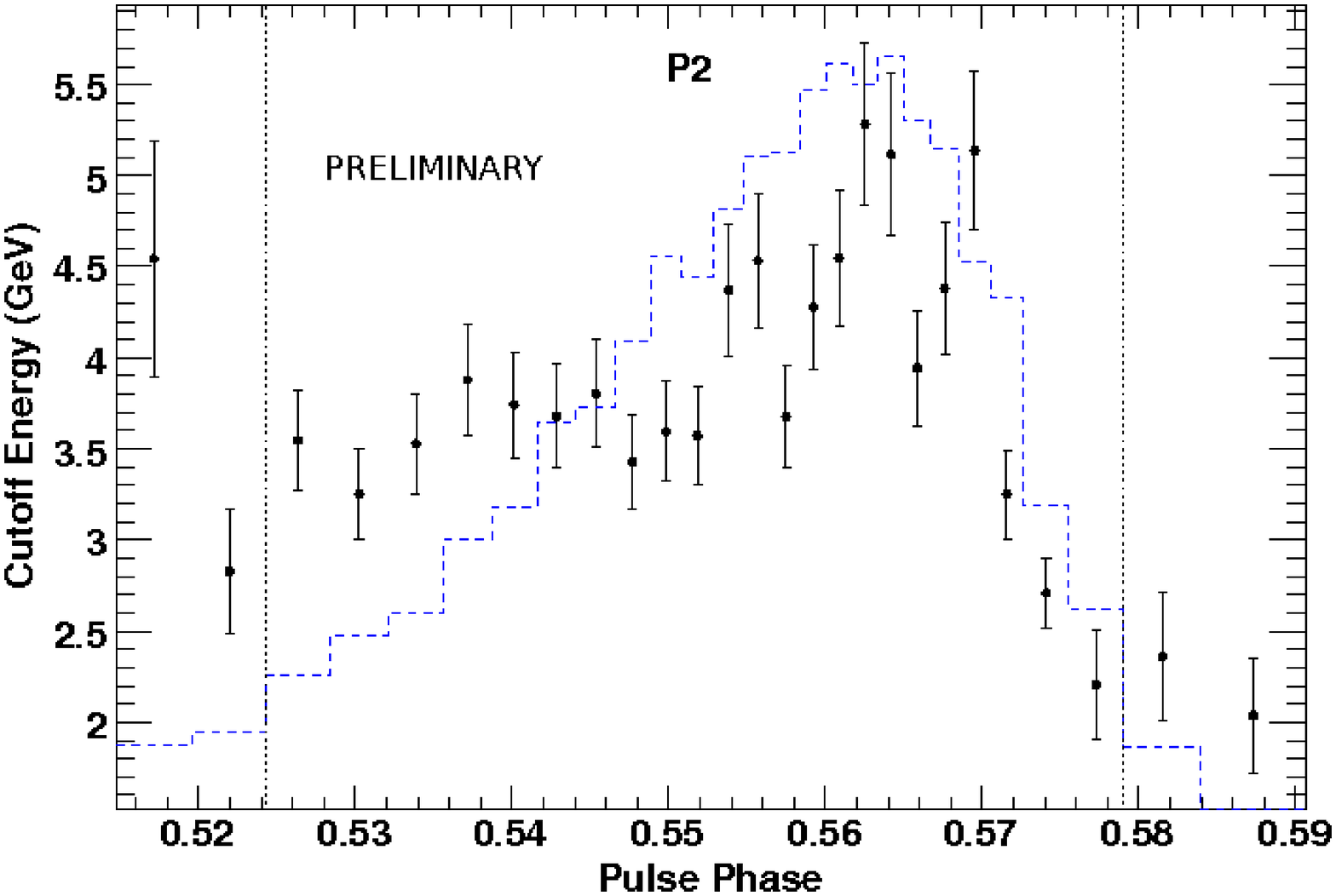}{0pt}{0}{270}{192.9}{-18}{0}
\caption{Cutoff energy vs phase for the second peak.  The photon index was held fixed to the best fit value of $1.58\pm0.01$ between the dashed vertical lines which correspond to the phase range $0.524\leq\phi\leq0.579$.}
\end{figure}

\section{Discussion}
The HE \g-ray emission from pulsars has been theorized to be curvature radiation (CR) from electrons (or positrons) accelerated along the magnetic field lines by the parallel component of the electric field (E$_{\parallel}$) (Romani 1996 and Hirotani \& Shibata 1999).  The electrons will be CR reaction limited such that a steady-state Lorentz factor is maintained.  In outer-magnetospheric emission models (e.g. Muslimov \& Harding 2004, slot gap (SG); Zhang et al. 2004; and Hirotani 2008, outer gap (OG)) E$_{\parallel}$ depends on the value of the magnetic field at the light cylinder and the gap width.  These models give similar cutoff energies, Equation 2 (units of mc$^{2}$), ranging from 1 to 5 GeV, consistent with what is observed in Vela and other \g-ray pulsars.

\begin{equation}
\rm{E}_{CR} = \frac{3}{2} \frac{\lambar}{\rho_{c}} \gamma^{3}_{CR} = 0.32\lambda_{c}\Big(\frac{\rm{E}_{\parallel}}{e}\Big)^{\frac{3}{4}} \rho^{\frac{1}{2}}_{c}
\end{equation}

Interestingly, the cutoff energy (E$_{CR}$) depends on the local field line radius of curvature ($\rho_{c}$).  Emission across the pulse originates from different ranges of emission radii (Romani 1996 and Cheng et al. 2000 (OG) and two-pole caustic (TPC) Dyks \& Rudak 2003), implying that phase-resolved spectroscopy should map out the emission altitude.  Large variations of $\rho_{c}$ with phase are also expected in these models and mapping the minimum $\rho_{c}$, using basic geometric models, can produce trends similar to what is seen in Figure 4, but full radiation models will be needed to match all of the features.  As the LAT continues to accumulate more events from Vela it will be possible to map the HE \g-ray emission regions in more detail and better understand the radiative processes involved.

\acknowledgements
\emph{Acknowledgements}\\
The \Fermi{} LAT Collaboration acknowledges support from a number of agencies and institutes for both development and the operation of the LAT as well as scientific data analysis. These include NASA and DOE in the United States; CEA/Irfu and IN2P3/CNRS in France; ASI and INFN in Italy, MEXT, KEK, and JAXA in Japan; and the K.~A.~Wallenberg Foundation, the Swedish Research Council and the National Space Board in Sweden. Additional support from INAF in Italy and CNES in France for science analysis during the operations phase is also gratefully acknowledged.

%The \Fermi{} LAT Collaboration acknowledges generous support from a number of agencies and institutes that have supported both the development and the operation of the LAT as well as scientific data analysis.  These include the National Aeronautics and Space Administration and the Department of Energy in the United States; the Commissariat \`a l'Energie Atomique and the Centre National de la Recherche Scientifique / Institut National de Physique Nucl\'eaire et de Physique des Particules in France; the Agenzia Spaziale Italiana and the Istituto Nazionale di Fisica Nucleare in Italy; the Ministry of Education, Culture, Sports, Science and Technology (MEXT), High Energy Accelerator Research Organization (KEK) and Japan Aerospace Exploration Agency (JAXA) in Japan; and the K.~A. Wallenberg Foundation, the Swedish Research Council and the Swedish National Space Board in Sweden.

%Additional Support for science analysis during the operations phase is gratefully acknowledged from the Istituto Nazionale di Astrofisica in Italy and the Centre National d'\'Etudes Spatiales in France.

The Parkes radio telescope is part of the Australia Telescope which is funded by the Commonwealth Government for operation as a National Facility managed by CSIRO. We thank our colleagues for their assistance with the radio timing observations.

\end{document}